# Pressure densification of a simple liquid

R. Casalini and C.M. Roland

Naval Research Laboratory, Chemistry Division, Washington, DC  20375-53342

*(Aug 9, 2017)*

ABSTRACT The magnitude of the high frequency, static dielectric permittivity is used to determine the density of tetramethyl tetraphenyl trisiloxane, a non-associated glass-forming liquid, as a function of temperature and pressure. We demonstrate that the properties in the glassy state are affected by the pressure applied to the liquid during vitrification. This behavior is normal for hydrogen-bonded liquids and polymers, but unanticipated by models of simple liquids.

________________________________________________________________

One of the curiosities regarding studies of the glass transition is the overriding focus on the properties of the liquid, rather than those of the glass. The main reasons for this are the equilibrium nature of the liquid state and the experimental inaccessibility of structural relaxation times, $\tau$, below the glass transition temperature, $T_g$.  These are, of course, the properties that define the glass transition – the material falls out of equilibrium as $\tau_\alpha$ becomes very large. In response to this nonequilibrium structure, glass slowly reorganizes, a process known as physical aging [1,2,3,4,5,6]. Aging is an important technical aspect of glasses, affecting their stability and thus utility for many applications. A factor controlling the non-equilibrium structure is the condition of the liquid upon vitrification. For example, variation of the rate of cooling through $T_g$ can be used to produce glasses with varying departures from equilibrium, and thus varying stability [7,8,9,10]. Another method, employed herein, is the application of pressure to the supercooled liquid. The glass transition is pressure-dependent, so that pressure affords a means to control the properties of the glass, including its physical aging behavior. Besides the material engineering value, understanding how the glass structure depends on the pressure during its formation is also important in discerning the principal control parameters and ultimately solving the "glass transition problem" [11,12,13,14,15,16,17].

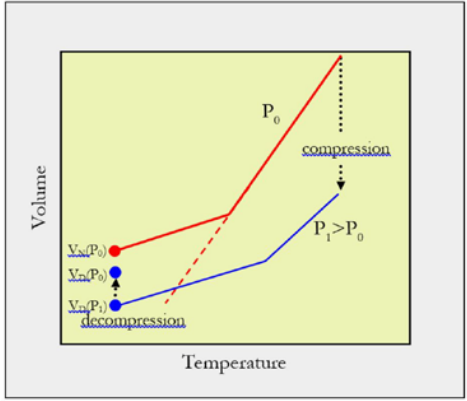

**Figure 1.** Pressure densification scheme: Glass formed by cooling at pressure $P_0$ has a lower density than when formed by cooling at higher pressure $P_1$, with the comparison made at the same temperature and pressure.

Inducing glass formation by supercooling a liquid under high pressure and then releasing the pressure after vitrification is known as pressure densification, the term a reference to the higher density of the glass, which forms at a higher temperature, in comparison to that obtained by preparing the glass in the conventional manner at low pressure. Most studies of pressure densification have been carried out on polymers [18,19,20,21,22,23,24] or compounds with ionic or hydrogen-bonding interactions [25,26,27,28]. Pressure densification studies of non-associated molecular liquids are scarce. Danilov et al. [29] showed that pressure-densified propylene carbonate had different properties, including higher modulus, than the corresponding glass prepared at ambient pressure. Pressure densification of simple liquids has special significance because of recent theoretical developments concerning their expected properties [30]. In the present context, simple liquids are defined as materials exhibiting (i) conformance to density scaling [31]

$$\tau = f\left(T\rho^{-\gamma}\right) \qquad (1)$$

in which $\rho$ is density, $\gamma$ a material constant, and $f$ a function; (ii) conformance to isochronal superpositioning [32]

$$\beta(T,P) = g(\tau) \qquad (2)$$

in which $\beta$ is the Kohlrausch stretch exponent describing the distribution of relaxation times, and $g$ is a function; and a Prigogine-Defay ratio not much larger than unity [33]. Such materials are limited to those in which the interactions are restricted to van der Waals forces and Coulombic forces, with no H-bonding or other strong associations nor a network structure [34].

Of interest herein is the prediction that simple liquids cannot be pressure densified; that is, the obtained glass is independent of the pressure applied during its formation [35,36,37]. However, literature data that enable assessment of this prediction are scarce. One reason for the lack of pressure densification experiments on molecular liquids is that most have very low glass transition temperatures. In a typical PVT apparatus using mercury as the confining liquid, the temperature range is limited by the freezing point of the mercury, 234.3K at atmospheric pressure, increasing 50 K per GPa [38]. `

In this work we avoid the problem by determining the density, $\rho$, indirectly from measurement of the dielectric constant (relative permittivity) of the liquid [39,40]. The Clausius–Mossotti equation relates the high frequency limiting value of the permittivity, $\varepsilon_\infty$, to the material density $\rho$

$$\frac{\varepsilon_\infty - 1}{\varepsilon_\infty + 2} = \frac{1}{3}\frac{\rho}{M}\frac{\alpha_0 N_A}{\varepsilon_0} \qquad (3)$$

where $\varepsilon_0$ is vacuum permittivity, $M$ the molecular weight, $N_A$ Avogadro's number, and $\alpha_0$ the sum of the atomic and electronic polarizabilities. Due to the very local nature of the induced polarization, in the investigated range $\alpha_0$ is essentially independent of pressure, eq. (3) reducing to

$$\frac{\varepsilon_\infty(T,P) - 1}{\varepsilon_\infty(T,P) + 2} = \varsigma(T)\rho(T,P) \qquad (4)$$

where the factor $\varsigma(T)$ depends only on temperature. To determine $\varepsilon_\infty$ from the dielectric spectra, the dielectric constant is measured at a frequency well beyond that of any absorption peaks.

The two thermodynamic pathways used to obtain the glassy state are represented in Figure 1, illustrating that glass formed at high pressure has a higher density than when formed at low pressure. A metric of this pressure densification is from the relative changes in specific volume (inverse density) [37]

$$\delta(P_0, P_1) = \frac{V_N(P_0) - V_D(P_0)}{V_N(P_0) - V_D(P_1)} \qquad (5)$$

Here $V_N$ and $V_D$ are the specific volumes for vitrification at low ($P_0$) and high ($P_1$) pressures, respectively. Substituting eq.(5) in eq.(4), $\varsigma(T)$ cancels out, and $\delta$ can be calculated from the ratio $\frac{\varepsilon_\infty + 2}{\varepsilon_\infty - 1}$.

The tetramethyl tetraphenyl trisiloxane (DC704 from Dow Corning) has a $T_g$ = 212K at atmospheric pressure, which is close to the minimum temperature attainable with our instrumentation. For this reason, $\delta$ was determined for $P_0$=158MPa and $T$=246.0±0.5K, which is 4K below $T_g$ at this pressure. For these conditions $\varepsilon'$ changes by less than 0.015% over 3 decades of frequency, and the dielectric loss is negligible, $\varepsilon''$ < $10^{-3}$. There are no secondary peaks in the spectra of DC704; thus, in applying eq. (4) we take $\varepsilon_\infty \cong \varepsilon'(10 kHz)$, measured at a temperature below $T_g$ for which the α-peak falls at frequencies more than 7 decades slower. The dielectric measurements were carried out with a Novocontrol Alpha Analyzer. The pressure vessel (Harwood Eng.) could apply up to 1.4 GPa hydrostatic pressure. Because any movement of the electrodes would alter the response, we employed an air capacitor (capacitance ~ 20 pF) immersed in the sample fluid and located inside a flexible Teflon cell. Thus, lateral changes in sample dimensions can occur but the thickness and geometrical capacitance are fixed. To verify no change in the geometrical capacitance during the experiments, after a measurement the pressure was decreased to 47MPa (T=246K), which brought the loss peak into the experimental window (see Figure 2). The change in the peak, and hence the change in geometrical capacitance, was less than 0.5%. Any adhesion of the sample to the metal capacitor could introduce small shear stresses; however, the bulk of the material experiences hydrostatic and uniform pressure.

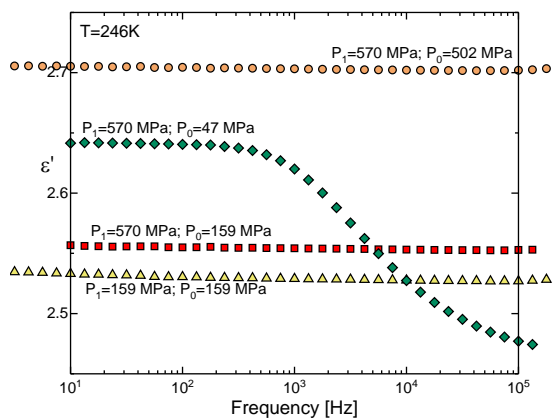

**Figure 2.** Real part of the permittivity measured for the pressure densified glass at high (circles) and low (squares) pressures, the glass formed at lower pressure (triangles), and the spectrum at a pressure sufficiently low that the relaxation falls within the measured frequency range (diamonds). The latter was used to ensure no disruption of the sample during changes in pressure.

Figure 2 shows representative dielectric measurements for the glass prepared at low and high pressures. Application of pressure to the liquid is followed by cooling below $T_g$, which causes contraction of the pressurizing fluid and a consequent small (*ca.* 10%) drop in pressure. From the dielectric constant data we calculate $\delta$ using eqs. (4) and (5). These are plotted in Figure 3.

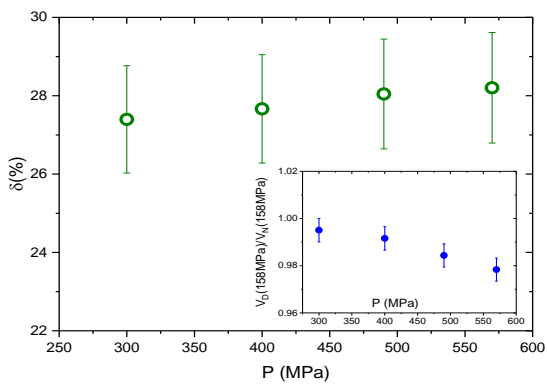

**Figure 3.** Pressure densification ratio $\delta$ (eq.(5)) versus densification pressure. (insert) Ratio of the specific volumes after cooling at two pressures, both measured at *T*=246K.

In the inset to Figure 3 is the relative change of specific volume as a function of the vitrification pressure. The change is about 2% at 570 MPa, comparable to that observed for propylene carbonate [29], but less than typical for pressure-densified polymers [18,19,20,21,22,23,25]. The magnitude of the pressure densification effect is substantial ($\delta \sim$ 27%), especially considering it is relative to 159 MPa, rather than the usual ambient pressure. The values of $\delta$ vary only weakly with pressure.

The fact that DC704 can be pressure densified at all is surprising, or at least at odds with the expectation that glasses formed from simple liquids have aging behavior unaffected by the conditions, including pressure, extant during the transition through $T_g$ [35,36,37]. As stated, the designation "simple" refers to liquids exhibiting certain dynamic properties. DC704 is the only material for which this panoply of properties defining simple liquids has been demonstrated experimentally [33,41,42]. Theoretical models [35,36] and molecular dynamics simulations [37] both indicate materials that have these properties cannot be pressure densified.

In summary, we utilize dielectric measurements of the static permittivity to characterize mass density changes in a glass-forming liquid under high pressure. The method extends the range of temperatures and pressures over which such information can be obtained. Applying this to a prototypical simple liquid, we find that when the pressure applied during formation of the glass is released, the consequent density is significantly greater than for a glass formed at lower pressure. This is normal behavior for polymers and associated liquids, but unexpected for simple liquids. Our finding suggests the need to re-examine the properties that define "simple" liquid behavior.

This work was supported by the Office of Naval Research.